\documentclass[journal]{IEEEtran}
\usepackage{amssymb,latexsym,graphicx,times,amsmath,subfigure,threeparttable,booktabs,bm,color,multirow,cite,amsmath,amsfonts,amsthm,bm}
\usepackage{color}
\makeatletter

\newcommand{\Rmnum}[1]{\expandafter\@slowromancap\romannumeral #1@}
\makeatother

\begin{document}
\title{Adversarial Artifact Detection\\ in EEG-Based Brain-Computer Interfaces}

\author{\IEEEauthorblockN{Xiaoqing Chen, Dongrui Wu}

\IEEEauthorblockA{School of Artificial Intelligence and Automation\\
Huazhong University of Science and Technology, Wuhan, China\\
Email: m202273198@hust.edu.cn, drwu@hust.edu.cn}}

\markboth{}
{Chen \MakeLowercase{\textit{et al.}}:Adversarial Artifact Detection in EEG-Based Brain-Computer-Interfaces}
\maketitle

\begin{abstract}

Machine learning has achieved great success in electroencephalogram (EEG) based brain-computer interfaces (BCIs). Most existing BCI research focused on improving its accuracy, but few had considered its security. Recent studies, however, have shown that EEG-based BCIs are vulnerable to adversarial attacks, where small perturbations added to the input can cause misclassification. Detection of adversarial examples is crucial to both the understanding of this phenomenon and the defense. This paper, for the first time, explores adversarial detection in EEG-based BCIs. Experiments on two EEG datasets using three convolutional neural networks were performed to verify the performances of multiple detection approaches. We showed that both white-box and black-box attacks can be detected, and the former are easier to detect.

\end{abstract}


\IEEEpeerreviewmaketitle

\section{Introduction}

\begin{figure}[htpb]\centering
{\includegraphics[width=1\linewidth,clip]{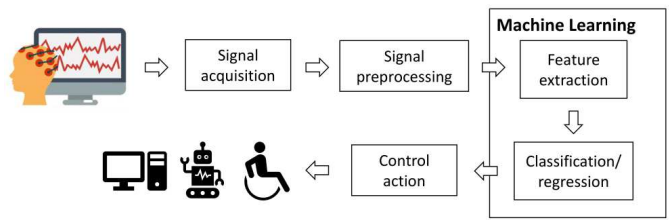}}
\caption{The flowchart of a BCI system. Manual feature extraction is necessary if traditional machine learning algorithms are employed.}
\label{fig:frame}
\end{figure}

A brain-computer interface (BCI), which has been extensively studied in neuroscience, neural engineering and clinical rehabilitation, builds a communication pathway between the human brain and a computer\cite{ienca2018brain}. Electroencephalogram (EEG), which records the brain's electrical activities from the scalp, has become the most widely used input signal in BCIs due to its low cost and convenience\cite{nicolas2012brain}. An EEG-based BCI system usually consists of four parts, namely signal acquisition, signal preprocessing, machine learning, and control action, as shown in Fig.~\ref{fig:frame}. The machine learning block includes feature extraction and classification/regression if traditional machine learning algorithms are adopted.

Machine learning has achieved great success in many applications \cite{he2016deep, devlin2018bert, parkhi2015deep}, including EEG-based BCIs. Despite their outstanding performance and robustness to random noise, machine learning models, especially deep learning models, are vulnerable to adversarial attacks, where carefully crafted human-imperceptible perturbations are added to benign samples to cause mis-recognitions \cite{szegedy2013intriguing, goodfellow2014explaining, brown2017adversarial}. The existence of adversarial examples raised wide attention and serious security concern about using machine learning models in safety/privacy-critical applications, such as malware detection \cite{grosse2016adversarial}, speech recognition \cite{carlini2018audio}, autonomous driving \cite{bar2020vulnerability}, etc.

For BCIs, most studies so far focused on increasing the accuracy and efficiency of machine learning algorithms, but few considered their security. However, as first revealed by Zhang and Wu \cite{zhang2019vulnerability}, adversarial examples generated by unsupervised fast gradient sign method (FGSM) \cite{goodfellow2014explaining} can significantly degrade the performance of deep learning classifiers in EEG-based BCIs \cite{wu2021adversarial}. Meng \emph{et al.} \cite{meng2019white} further exposed the vulnerability of machine learning algorithms in regression tasks of EEG-based BCIs. Zhang \emph{et al.} \cite{zhang2021tiny} showed that adversarial examples can fool BCI spellers to output any wrong character the attacker wants. Liu \emph{et al.} \cite{liu2021universal} designed a total loss minimization approach to craft universal adversarial perturbations for EEG-based BCIs, making adversarial attacks easier to implement. Bian \emph{et al.} \cite{bian2022ssvep} used simple square wave signals to generate adversarial examples, which can mislead steady-state visual evoked potential based BCIs. Jiang \emph{et al.} \cite{drwuAP2022} used active learning for efficient poisoning attacks to EEG-based BCIs.

The consequences of adversarial attacks to BCIs could range from merely user frustration to severe injury, raising a critical safety concern and an urgent need for adversarial defense. For example, adversarial attacks can cause malfunctions in exoskeletons or wheelchairs controlled by EEG-based BCIs \cite{li2015multimodal}, and may drive the user into danger on purpose. In BCI-based driver drowsiness estimation \cite{wu2016driver}, adversarial attacks may hijack the output of the BCI system and increase the possibility of traffic accidents. More importantly, in military settings, adversarial attacks in BCIs may result in false commands, such as friendly fire \cite{RR-2996-RC}. Therefore, it is critical to develop adversarial defense approaches for BCIs.

In the literature, multiple adversarial defense approaches have been proposed for other applications, e.g., computer vision and natural language processing, which can be divided into proactive defenses and reactive defenses \cite{aldahdooh2022adversarial}. Proactive defenses, such as adversarial training \cite{zhang2019theoretically, madry2017towards}, aim at obtaining a robust classifier. Reactive defenses \cite{feinman2017detecting, ma2018characterizing, lee2018simple} attempt to identify adversarial examples to reject them or restore them to normal ones.

Though various approaches have been proposed to tackle adversarial attacks, challenges still exist \cite{carlini2017adversarial, deng2021libre}. In adversarial training approaches (proactive defense), the inclusion of adversarial examples into training can not only incur extra training overheads but also compromise the model's performance on benign examples, and classifiers trained to accommodate a certain attack may later be defeated by new attack strategies. Aimed at exploring the intrinsic properties of adversarial examples, adversarial detection approaches (reactive defense) based on designed statistics have shown satisfactory robustness to known optimization-based attacks and generalization ability toward unseen attacks. However, they are usually task-specific, e.g., for image classification only \cite{feinman2017detecting, ma2018characterizing, lee2018simple, zhang2018detecting}. A given detection approach performing well for one task may perform poorly for another. To our knowledge, adversarial detection in EEG-based BCIs has not been studied yet.

This paper implements and compares several state-of-the-art adversarial detection approaches for EEG-based BCIs. Experiments on two EEG datasets and three convolutional neural network (CNN) models showed that by extracting features from the output of neural networks, we can effectively distinguish between white-box adversarial examples and normal ones.

The remainder of this paper is organized as follows: Section~\ref{sect:TM} introduces several adversarial attacks used in our experiments. Section~\ref{sect:DA} describes our adversarial detection strategies. Section~\ref{sect:ES} details our performance evaluation settings. Section~\ref{sect:daa} presents the detection results. Section~\ref{sect:CFR} draws conclusions and points out some future research directions.

\section{Adversarial Attacks} \label{sect:TM}

Let $\mathbf{\emph{C}_{\bm{\theta}}}(X):\mathcal{X}\rightarrow \mathcal{Y}$ be an EEG classifier, where $\bm{\theta}$ is model parameters, $X\in \mathcal{X} \subset \mathbb{R}^{C\times T}$ is an EEG epoch where $C$ stands for the number of EEG channels and $T$ the number of time domain samples, and $\mathcal{Y}=\{1,2,...,k\}$, with $k$ being the number of classes. Let $X^{adv}$ be an adversarial example generated from $X$, and $y$ the ground-truth label for $X$. In adversarial attacks, $X^{adv}$ needs to satisfy:
\begin{align}
\mathbf{\emph{C}_{\bm{\theta}}}(X^{adv})\neq y,\\
\emph{D}(X^{adv},X)\textless {\epsilon},
\end{align}
where $\emph{D}$ is a given distance metric, and ${\epsilon}$ is a predefined perturbation threshold. A common choice of $\emph{D}$ is $\ell_p$ norm, and $\ell_\infty$ is used in this paper.

To satisfy the above constraints, $X^{adv}$ can be re-expressed as \cite{madry2017towards}:
\begin{align}
X^{adv}=X+ \mathop{\arg\max}_{\|\bm{\delta}\|\textless{\epsilon}}\mathcal{L}(\mathbf{\emph{C}_{\bm{\theta}}}(X+\bm{\delta}),y),
\end{align}
where $\mathcal{L}$ is the classifier's loss function, $\|\cdot\|$ stands for $\ell_p$ norm, and $\bm{\delta}$ is the calculated adversarial perturbation.

According to how much the attacker knows about the target model, adversarial attacks can be categorized into white-box attacks and black-box attacks. White-box attacks assume the adversary has full access to the model architecture and parameters, whereas in black-box attacks, the adversary can only acquire the model's predicted classes or probabilities. Most white-box attacks are designed based on optimization or gradient strategies, and black-box attacks are implemented based on the transferability between models or by querying the model outputs. Black-box attacks are more challenging to implement than white-box attacks due to less knowledge acquired, but are more practical.

To fully explore the characteristics of different attacks, this paper implements two gradient-based white box attacks, i.e., FGSM \cite{goodfellow2014explaining} and projected gradient descent (PGD) \cite{madry2017towards}, one optimization-based white-box attack, i.e., Carlini-Wagner (CW) attack \cite{carlini2017towards}, and one transferability-based black-box attack, to EEG-based BCIs.

\subsection{Fast Gradient Sign Method (FGSM)}

FGSM \cite{goodfellow2014explaining}, a simple yet powerful attack approach, crafts the adversarial example via one-step gradient calculation:
\begin{align}
X^{adv}=X+{\epsilon}\cdot {\rm sign}(\nabla_{X}\mathcal{L}(\mathbf{\emph{C}_{\bm{\theta}}}(X),y)).\label{fgsm}
\end{align}
Along the direction of the gradient, the loss of the model output with respect to the ground-truth label is increased, forcing the model to misclassify. Increasing $\bm{\epsilon}$ usually improves the attack success rate.

\subsection{Projected Gradient Descent (PGD)}

PGD \cite{madry2017towards} applies FGSM multiple times with a smaller step size, and bounds the magnitude of the generated adversarial perturbation after each iteration.

More specifically, PGD starts from a random point near the benign example $X$:
\begin{align}
X_0^{adv} = X+ \bm{\xi},
\end{align}
where $\bm{\xi}\in\mathcal{U}(-\epsilon,\epsilon)$ is random noise, and then iterates:
\begin{align}
X_i^{adv} = {\rm Proj}_{X,\epsilon}\left(X_{i-1}+{\alpha}\cdot{\rm sign}(\nabla_{X_{i-1}}\mathcal{L}(\mathbf{\emph{C}_{\bm{\theta}}}
(X_{i-1}),y))\right),\label{pgd}
\end{align}
where $\alpha\le \epsilon$ is the attack step size, and $i=1,2,...,n_{iter}$, with $n_{iter}$ being the number of iterations. ${\rm Proj}_{X,\epsilon}$ clips the value of the input to keep $X_i^{adv}$ in the $\epsilon$ neighborhood of the benign example $X$ under $\ell_\infty$ norm.

\subsection{Carlini-Wagner (CW) Attack}

CW attack was proposed by Carlini and Wagner \cite{carlini2017towards}, which is now one of the strongest white-box attacks. Carlini and Wagner formulated adversarial attack as the following optimization problem:
\begin{align}\label{cw}
\min_{\bm{\delta}} \|\bm{\delta}\|_2+ c\cdot h(X+\bm{\delta}),
\end{align}
where $c$ is a trade-off parameter found by binary search during the optimization process, and $h(\cdot)$ is designed in such a way that $h(X+\bm{\delta})\le 0$ if and only if $\emph{C}_{\bm{\theta}}(X+\bm{\delta})\neq y$. Carlini and Wagner designed $h(X')$ as a hinge loss:
\begin{align}
h(X')=\max\left(\mathbf{\emph{Z}}(X')_y-\max_{i\neq y}\mathbf{\emph{Z}}(X')_i,-\kappa\right),
\end{align}
where $\mathbf{\emph{Z}}(X')$ is the output of the classifier's penultimate layer, and $\kappa$ is a hyperparameter called confidence. An adversarial example with a higher $\kappa$ usually has larger perturbation size and better transferability.

For a given example $X$, CW attack tries to find a perturbation $\bm{\delta}$ that is small in size ($\|\cdot\|$ in the loss function) but can mislead the classifier [$h(\cdot)$ in the loss function].

After optimization, we clip the generated adversarial perturbation so that the adversarial examples are kept in the ${\epsilon}$ neighborhood of the benign example.

\subsection{Transferability-based Black-box Attack}

In black-box attacks, the adversary can only obtain the model output for an input. Black-box attacks are generally more challenging than white-box attacks because of less knowledge acquired, but they are more practical.

Transferability-based black-box attacks leverage the transferability of adversarial examples between models to attack the target model \cite{papernot2016practical}, i.e., for a certain task, adversarial examples generated using one model can confuse other models with high probability, regardless of the differences of model architectures and parameters.

As in Zhang and Wu's work \cite{zhang2019vulnerability}, first, a substitute model is trained utilizing generated data labeled with the target model's outputs; then, adversarial examples crafted using the substitute model via white-box attack strategies are used to attack the target model. We used FGSM in black-box attacks, as in \cite{zhang2019vulnerability}.

\section{Adversarial Detection} \label{sect:DA}

Although BCIs are vulnerable to adversarial attacks, adversarial detection in BCIs has not been explored yet.

\subsection{Bayesian Uncertainty}

Adversarial detection based on kernel density and Bayesian uncertainty (BU) has been shown as the most robust against optimization-based white-box attacks among ten different adversarial defense approaches \cite{carlini2017adversarial}, including adversarial retraining and input transformation.

A deep neural network trained with dropout is an approximation of the deep Gaussian process \cite{gal2016dropout}, a type of Bayesian model, whose uncertain output can be used to estimate low-confidence regions of the input space where adversarial examples may lie. Given a set of functions $\mathcal{G}$ that can map the input space to the output space, the prediction of each function $g \in \mathcal{G} $ is computed, and in a Gaussian process, the variance of the output values can be used as an indicator of the model's uncertainty \cite{feinman2017detecting}.

For a test example $X$ and its stochastic prediction scores $\{\mathbf s_1,...,\mathbf s_T\}$ generated by a neural network with dropout enabled, the BU of the model can be computed as \cite{feinman2017detecting}:
\begin{align}
{\rm BU}(X)=\frac{1}{T}\sum_{i=1}^T {\mathbf s_i}^\top \mathbf s_i-\left(\frac{1}{T}\sum_{i=1}^T {\mathbf s_i}\right)^\top \left(\frac{1}{T}\sum_{i=1}^T {\mathbf s_i}\right).
\end{align}

Feinman \emph{et al.} \cite{feinman2017detecting} claimed that the BU for an adversarial example $X^{adv}$ falling near (but not on) the benign data manifold is high. Research has also found that models can be very confident about some adversarial examples  \cite{goodfellow2014explaining}, which Feinman \emph{et al.} \cite{feinman2017detecting} assumed to lie far away from the benign data manifold and can be detected via distance-based metrics.

\subsection{Local Intrinsic Dimensionality}

To better understand the properties of adversarial examples, Ma \emph{et al.} \cite{ma2018characterizing} proposed to use local intrinsic dimensionality (LID) to characterize adversarial regions. LID evaluates the space-filling ability of the region surrounding a reference example via the distance distribution of its neighbouring examples, which can reflect the intrinsic data dimensionality \cite{houle2017local}. They argued that the LID of an adversarial example should be far higher than that of a benign example.

Given an example $X \in \mathcal{X}$, its LID is calculated as \cite{ma2018characterizing}:
\begin{align}
{\rm LID} (X)=-\left(\frac{1}{k}\sum_{i=1}^k {\rm log} \frac{r_i(X)}{r_k(X)}\right)^{-1},
\end{align}
where $r_i(X)$ denotes the distance between $X$ and its $i$th nearest neighbor among a batch of samples drawn from $\mathcal{X}$, and $k$ is the number of neighbors taken into account. To perform more efficient calculations, Ma \emph{et al.} \cite{ma2018characterizing} calculated LID from a randomly selected batch of examples from the dataset. While a larger batch of data can often lead to a more precise estimation of LID, they showed that discrimination between adversarial and benign examples can be achieved using a minibatch size as small as 100 and $k$ as small as 20.

\subsection{Mahalanobis Distance-based Confidence Score}

Lee \emph{et al.} \cite{lee2018simple} proposed a Mahalanobis distance-based metric for the detection of adversarial examples. Given pre-trained features of a softmax neural classifier $f(X)$, they computed the empirical class mean $\widehat{\bm{\mu}}_c$ and covariance $\widehat{\bm\sigma}$ of training samples $\{(X_1,y_1),...,(X_N,y_N)\}$ as:
\begin{align}
\widehat{\bm{\mu}}_c&=\frac{1}{N_c}\sum_{i:y_i=c}f(X_i),\\
\widehat{\bm{\sigma}}&=\frac{1}{N}\sum_{c}\sum_{i:y_i=c}(f(X_i)-\widehat{\bm{\mu}}_c)(f(X_i)-\widehat{\bm{\mu}}_c)^T,
\end{align}
where $N_c$ is the number of samples belonging to Class $c$. Using the above induced class mean $\widehat{\bm{\mu}}_c$ and covariance $\widehat{\bm{\sigma}}$, they defined the Mahalanobis distance-based confidence score ${\rm MD}_{\max}(X)$:
\begin{align}
{\rm MD}_{\max}(X)= \max_{c}-(f(X)-\widehat{\bm{\mu}}_c)^T\widehat{\bm{\sigma}}^{-1}(f(X)-\widehat{\bm{\mu}}_c).\label{mx}
\end{align}

Lee \emph{et al.} \cite{lee2018simple} further proposed an input calibration technique to better distinguish adversarial examples from benign ones. For each test sample $X$, they calculated the calibrated $\widehat{X}$ by adding small noise to $X$:
\begin{align}
\widehat{X}&=X+{\varepsilon} \cdot{\rm sign}(\nabla_{X}{\rm MD}_{\max}(X))\\
&=X-{\varepsilon}\cdot {\rm sign}((f(X)-\widehat{\bm{\mu}}_c)^T\widehat{\bm{\sigma}}^{-1}(f(X)-\widehat{\bm{\mu}}_c)),
\end{align}
where $c$ is the closest class to $X$ under the Mahalanobis distance, which can be identified during the calculation of ${\rm MD}_{\max}(X)$. After calibration, $\widehat{X}$ is closer to the class mean vector  $\widehat{\bm{\mu}}_c$ than  $X$ based on the Mahalanobis distance.

\section{Experimental Setup} \label{sect:ES}

\subsection{Datasets}

We conducted experiments using the following two datasets:
\begin{enumerate}
    \item Feedback error-related negativity (ERN) \cite{margaux2012objective}: The ERN dataset was released in a Kaggle challenge$\footnote{https://www.kaggle.com/competitions/inria-bci-challenge}$ at the 2015 IEEE Neural Engineering Conference. This dataset was collected from 26 subjects for two-class classification (good-feedback and bad-feedback) tasks. It consists of a training set of 16 subjects and a test set of 10 subjects. We only used the training set in our experiments. We downsampled the 56-channel EEG data to 200 Hz and filtered them using a [1,40] Hz band-pass filter. EEG epoches between [0,1.3]s were extracted and z-score standardized for later tasks, and each subject had 340 EEG epochs.
    \item Motor imagery (MI) \cite{tangermann2012review}: The MI dataset is Dataset 2A in BCI Competition IV$\footnote{https://www.bbci.de/competition/iv/}$. The dataset was collected from 9 subjects for four-class classification tasks (left hand, right hand, feet and tongue). The 22-channel EEG signals were sampled at 128 Hz and were filtered by our [4,40] Hz band-pass filter. We extracted the data in [0,2]s after each imagination prompt and standardized them using an exponential moving average window with a decay factor of 0.999. Each subject had 144 EEG epochs for every class.
\end{enumerate}

\subsection{Evaluated Models}

The following three CNN models were used in our experiments:
\begin{enumerate}
	\item EEGNet \cite{lawhern2018eegnet}: EEGNet is a compact CNN model custom-made for EEG classification tasks, which contains two convolutional blocks and one classification block. Depthwise and separable convolutions are used in EEGNet instead of traditional ones, which reduces the number of model parameters.
	\item DeepCNN \cite{schirrmeister2017deep}: DeepCNN is larger in size than EEGNet. It contains four convolutional blocks and a softmax layer for classification. The first convolutional block is specially designed for EEG inputs and the other three are standard ones.
	\item ShallowCNN \cite{schirrmeister2017deep}: Inspired by filter bank common spatial patterns, ShallowCNN is a shallow version of DeepCNN. It has one convolutional block with a larger kernel, a different activation function and a different pooling approach compared with DeepCNN.
\end{enumerate}

\subsection{Evaluation Settings}

We conducted leave-one-subject-out cross-validation on each dataset. For a dataset with $N$ subjects, one subject was used as the test set, two subjects were used to train two substitute models respectively if transferability-based black-box attack was performed, or otherwise discarded, and the remaining subjects were used to train the target model. In our transferability-based black-box attack, the first substitute model was used to imitate the substitute model trained by the adversary, and the second to craft adversarial examples for training our adversarial detectors.

All adversarial detectors were trained on the training set with equal numbers of normal and adversarial examples. We extracted features from the last layer of neural networks, which was found the most effective in \cite{ma2018characterizing}. In white-box attack detection, we crafted adversarial examples directly on the target model. In black-box attack  detection, we trained our adversarial detectors using adversarial examples generated on the second substitute model. Benign examples which can be correctly classified by the target model and whose corresponding adversarial examples can successfully fool the target model were used in the white-box attack experiments and the training of our black-box adversarial example detectors. In the testing phase of black-box adversarial example detection, we used all benign examples that can be correctly classified by the target model, regardless of whether their corresponding adversarial examples were effective or not. Because black-box attacks are generally weaker than white-box attacks, the number of successful adversarial examples was smaller.

We crafted FGSM, PGD, CW and black-box adversarial examples under $\ell_\infty$ norm constraint of 0.1. We set the perturbation amplitude ${\epsilon}=0.1$ in FGSM and PGD, the perturbation step size ${ \alpha}=0.01$ in PGD, and its number of iterations to 30. In CW attack, we clipped the perturbation value to [-0.1,0.1].

\subsection{Parameter Tuning}

The number of neighbors $k$ for LID and the noise magnitude $\bm{\varepsilon}$ for the Mahalanobis distance-based detector were chosen using nested cross-validation within the training dataset, based on the AUC values of the detection ROC curve. For LID, the number of nearest neighbors was tuned using grid search in [50, 90] with mini-batch size 100, as in \cite{ma2018characterizing}. For the Mahalanobis method, we tuned $\bm{\varepsilon}$ using an exhaustive grid search in the log-space of [$1E^{-5}$,$1E^{-2}$].

\subsection{Performance Measures}

We used both raw classification accuracy (RCA) and balanced classification accuracy (BCA) to evaluate the classification performance of the models. The RCA is the ratio of the number of correctly classified examples to the number of total examples. The BCA is the average of the RCAs of different classes. BCA is necessary because the ERN dataset has significant class imbalance, which is the case of real-world EEG classifications; using RCA alone can be misleading here.

For adversarial detectors, we used the AUC score to evaluate their performance. The AUC score is the area under the receive operating characteristic curve and ranges from 0 to 1. In a binary classification task, using the AUC score to measure the performance of the classifier avoids the error caused by the selection of classification threshold. The closer the AUC value of a classifier is to 1, the better the performance of this classifier.

\section{Experimental Results} \label{sect:daa}

\subsection{Attack Performance}

Knowing the performances and characteristics of various attacks helps understand adversarial examples. Before adversarial detection, we computed the BCAs and RCAs of neural network classifiers on various adversarial examples, and the $\ell_2$ norms of adversarial perturbations imposed by various adversarial attacks.

The results of white-box attacks and the corresponding $\ell_2$ norms of the adversarial perturbations are shown in Table~\ref{tab:white-att}. Adversarial examples created by FGSM, PGD and CW attacks were almost all successful in deceiving the classifiers. $\ell_2$ norms of the adversarial perturbation imposed by PGD were slightly smaller than those by FGSM, and those by CW attack were significantly smaller.

\begin{table*}[htbp] \centering \setlength{\tabcolsep}{1.2mm}
\caption{RCAs/BCAs of different CNN classifiers under FGSM, PGD and CW attacks, and $\ell_2$ norms of their corresponding adversarial perturbations.}   \label{tab:white-att}
\begin{tabular}{c|c|cccc|ccc} \toprule
\multirow{2}{*}{Dataset}&\multirow{2}{*}{Model} &\multicolumn{4}{|c}{RCAs/BCAs} &\multicolumn{3}{|c}{$\ell_2$ Norm} \\\cline{3-9}
&&Baseline&FGSM &PGD &CW&\makebox[0.05\textwidth][c]{FGSM} &\makebox[0.05\textwidth][c]{PGD} &\makebox[0.05\textwidth][c]{CW}\\  \midrule
\multirow{3}{*}{ERN}&
EEGNet &.6585/.6472 &.0001/.0005	&.0001/.0005	&.0001/.0005&12.07 &11.87	&1.16\\
&DeepCNN & .6647/.6359	&.0013/.0019	&.0001/.0005	&.0001/.0005	& 11.88	&11.02	&1.31\\
&ShallowCNN &.6818/.6312&	.0001/.0003&	.0001/.0003	&.0000/.0000&12.00&10.55&	1.37\\ \midrule
\multirow{3}{*}{MI}&EEGNet& .4515/.4515&.0000/.0000&	.0000/.0000	&.0000/.0000& 7.50&6.75&	0.39\\
&DeepCNN & .4418/.4418	&.0001/.0001	&.0000/.0000	&.0000/.0000	&7.45	&7.07	&0.92\\
&ShallowCNN & .4511/.4511	&.0000/.0000&.0000/.0000&	.0041/.0041&7.42	&6.79&0.77	\\ \bottomrule
\end{tabular}
\end{table*}

The rank of $\ell_2$ norms (magnitudes of the perturbations) is intuitive. FGSM is a simple one-step approach. PGD uses a smaller step size to iteratively perform the attack. Compared with FGSM, PGD is more sophisticated and therefore better able to find smaller adversarial perturbations. CW performs multiple optimizations to find smaller adversarial perturbations with guaranteed attack success, which largely reduces the magnitude of the imposed adversarial perturbations.

\begin{figure}
\centering
\subfigure[]{
\begin{minipage}[b]{0.4\textwidth}
\includegraphics[width=1\textwidth]{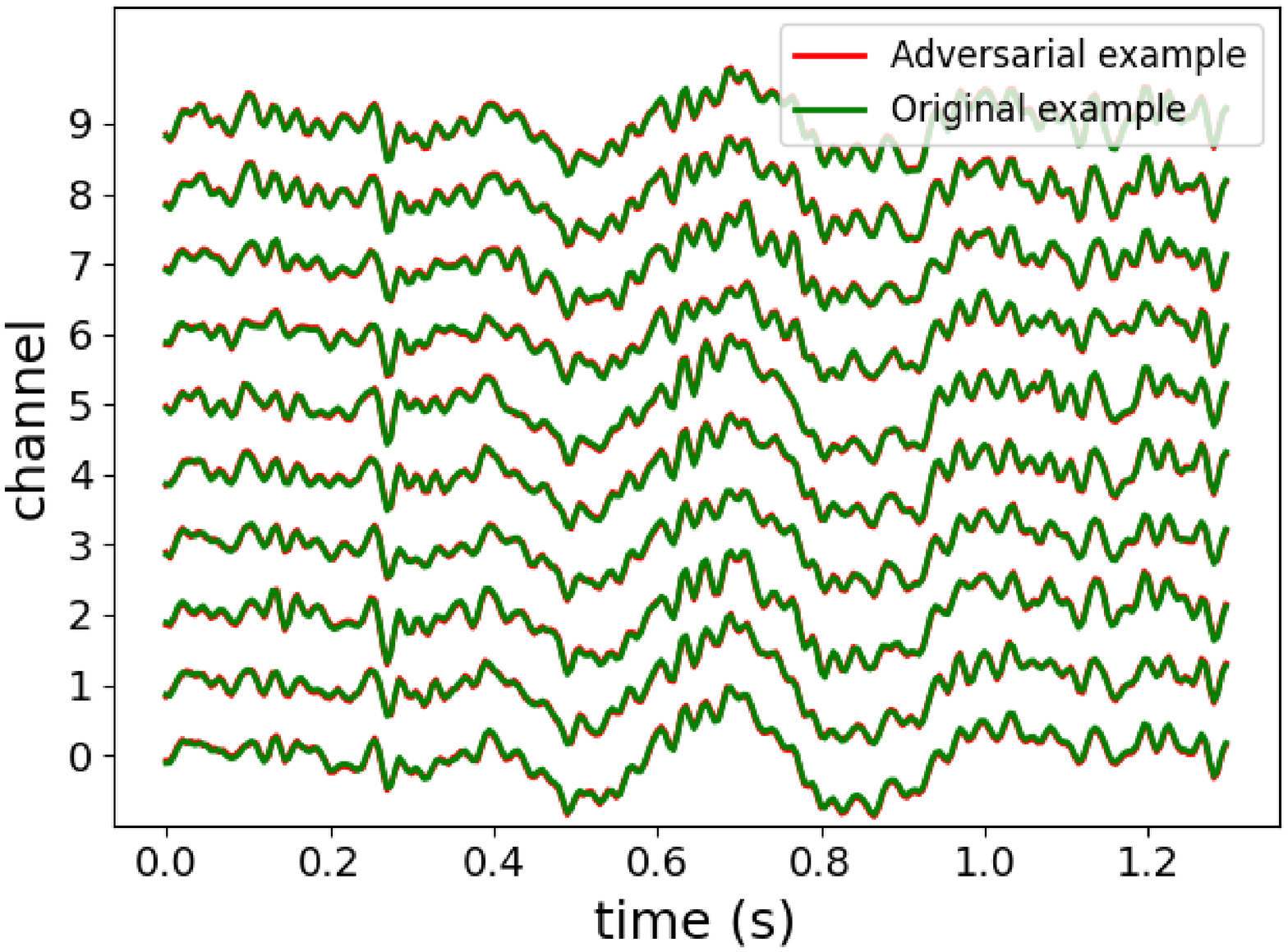}\vspace{0mm}
\end{minipage}
}\\
\subfigure[]{
\begin{minipage}[b]{0.4\textwidth}
\includegraphics[width=1\textwidth]{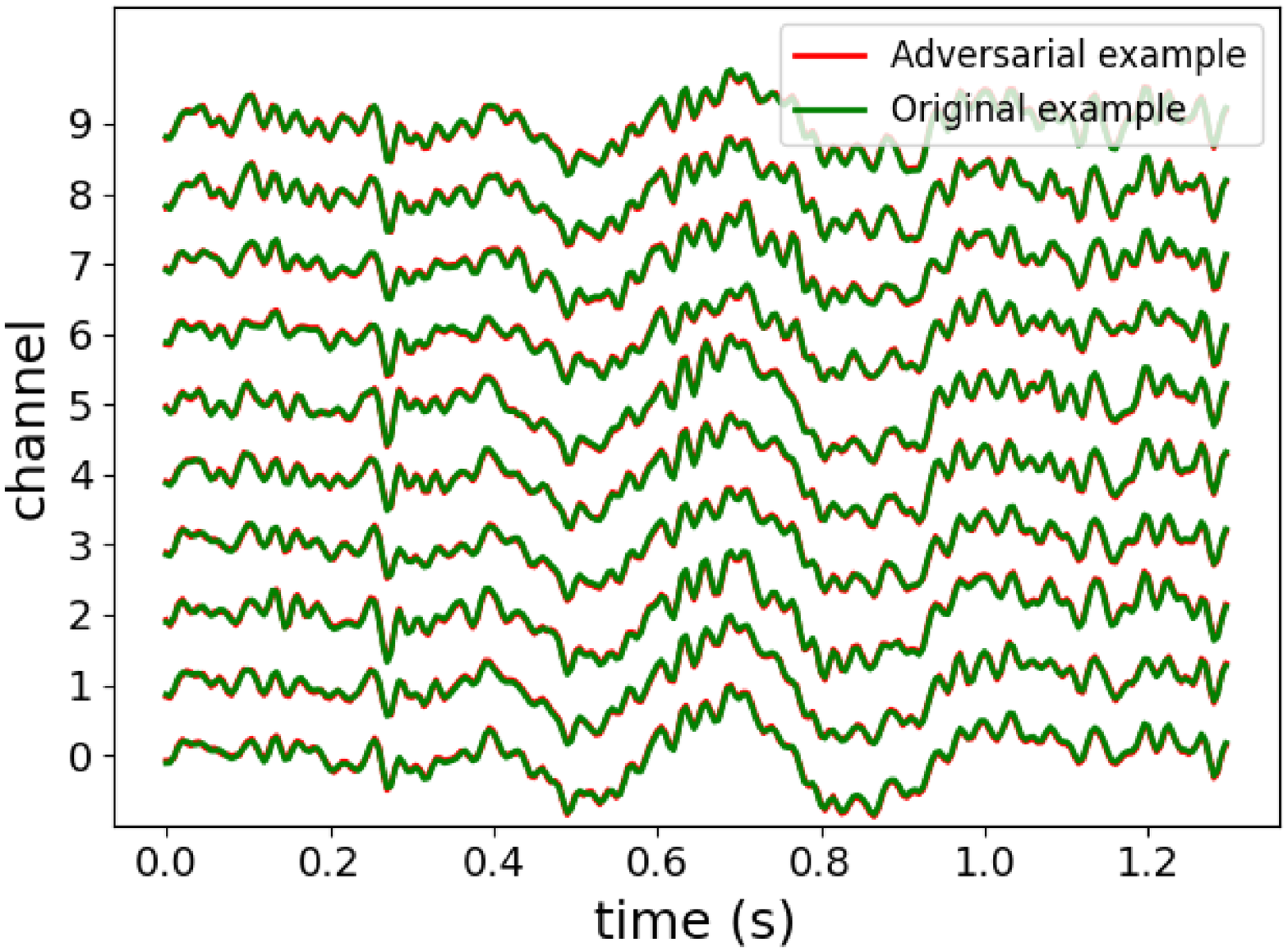}
\end{minipage}}
\\
\subfigure[]{
\begin{minipage}[b]{0.4\textwidth}
\includegraphics[width=1\textwidth]{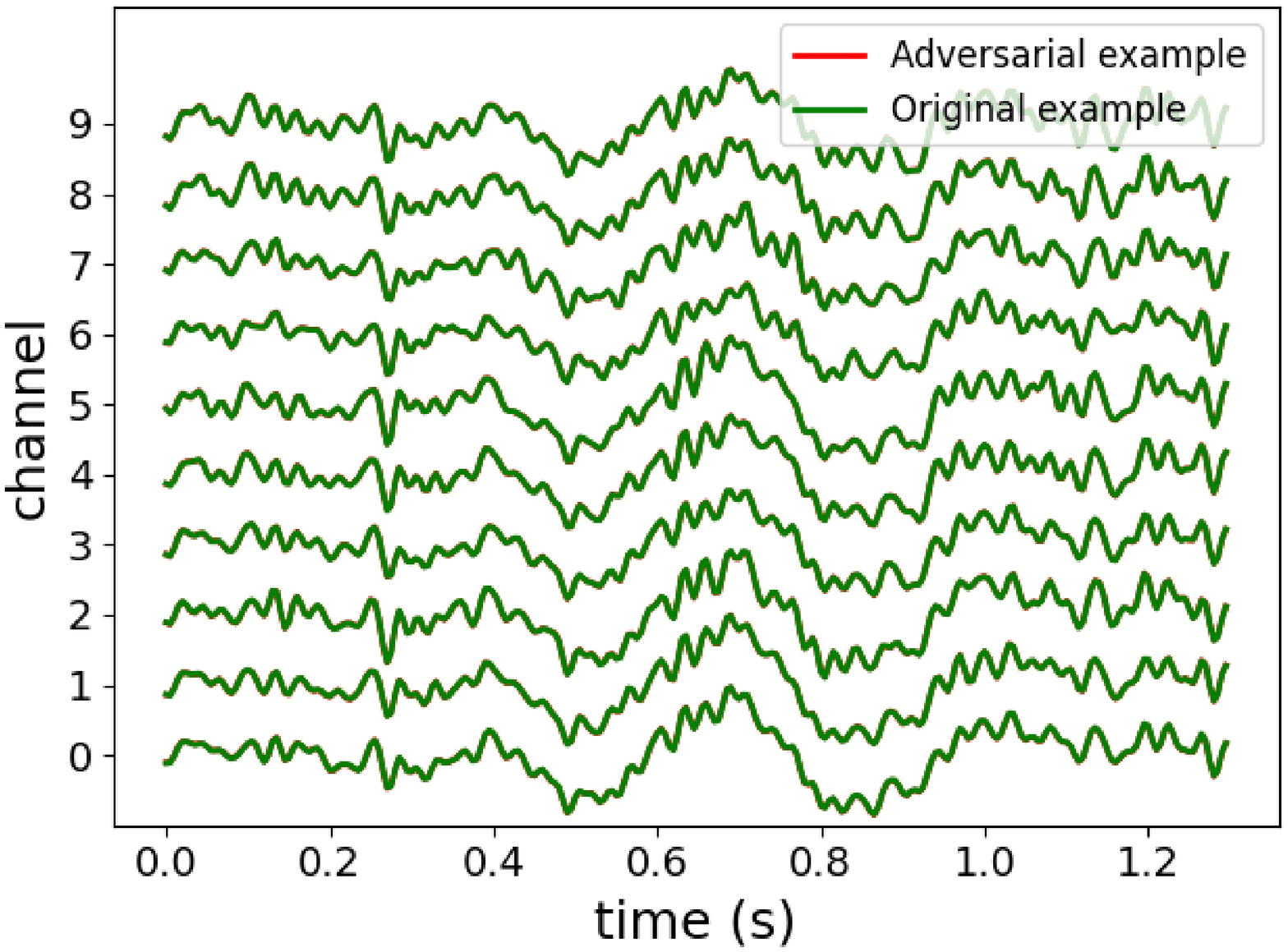}
\end{minipage}
}\caption{Examples of a P300 EEG epoch before and after adversarial
attack. (a) FGSM; (b) PGD; and (c) CW.}\label{attacks}
\end{figure}

Examples of a P300 EEG epoch before and after adversarial attacks are shown in Fig.~\ref{attacks}. All three adversarial perturbations are tiny and difficult to be noticed by human eyes.

We used EEGNet, DeepCNN and ShallowCNN as the substitute model to craft black-box adversarial examples, respectively. The results of FGSM black-box attacks and the corresponding $\ell_2$ norms of the adversarial perturbations are shown in Table~\ref{tab:black-att}. Black-box attacks can also greatly reduce the classification accuracy of the neural networks. In general, the attacks were more effective when the substitute model and the target model had the same structure. The $\ell_2$ norms in Table~\ref{tab:black-att} are similar to those of FGSM attacks in Table~\ref{tab:white-att}.

\begin{table*}[htbp] \centering \setlength{\tabcolsep}{1.2mm}
\caption{RCAs/BCAs of different CNN classifiers under FGSM black-box attack, and $\ell_2$ norms of their corresponding adversarial perturbations.}   \label{tab:black-att}
\begin{tabular}{c|c|cccc|ccc} \toprule
\multirow{2}{*}{Dataset}&\multirow{2}{*}{Target Model}&\multicolumn{4}{|c}{RCAs/BCAs}&\multicolumn{3}{|c}{$\ell_2$ norm}\\\cline{3-9}
&&Baseline&EEGNet &DeepCNN&ShallowCNN &\makebox[0.07\textwidth][c]{EEGNet} &\makebox[0.07\textwidth][c]{DeepCNN}&\makebox[0.07\textwidth][c]{ShallowCNN }\\  \midrule
\multirow{3}{*}{ERN}&
EEGNet &.6585/.6472 &.1352/.1416	&.1730/.1794&.5814/.5789&12.07 &11.88&12.00	\\
&DeepCNN & .6647/.6359	&.2754/.2686&.1767/.1711&.5995/.5701	& 12.07	&11.88	&12.00	\\
&ShallowCNN &.6818/.6312&	.5479/.5016&.5426/.4897&.3458/.2954	& 12.07	&11.88	&12.00\\ \midrule
\multirow{3}{*}{MI}&EEGNet& .4515/.4515& .0372/.0372&	.2587/.2587	&.1707/.1707& 7.50&7.45&	7.42\\
&DeepCNN & .4418/.4418	&.3568/.3568	&.1836/.1836&.1214/.1214	& 7.50&7.45&	7.42	\\
&ShallowCNN & .4511/.4511	&.3558/.3558&.2495/.2495&.0774/.0774	& 7.50&7.45&	7.42\\ \bottomrule
\end{tabular}
\end{table*}

\subsection{Detection of White-box Attacks} \label{sect:dw}

Table~\ref{tab:white-dec} shows the discrimination performance (AUC score) of three adversarial detectors (BU, LID, and Mahalanobis distance-based detector) on two EEG datasets and three EEG classifiers. We compared the adversarial detection AUC scores under two gradient-based adversarial attacks (FGSM and PGD) and one optimization-based attack (CW).

\begin{table}
\centering
\caption{AUC scores (\%) for different adversarial detection approaches under white-box attacks.}
\label{tab:white-dec}
\begin{tabular}{c|c|c|ccc}
\toprule
Attack       & Dataset              & Model      & \makebox[0.01\textwidth][c]  {BU}    & \makebox[0.01\textwidth][c] {LID}            & \makebox[0.01\textwidth][c] MD$_{\max}$  \\
\midrule
\multirow{6}{*}{FGSM} & \multirow{3}{*}{ERN} & EEGNet     & 91.75 & \textbf{96.49} & 93.62       \\
                      &                      & DeepCNN    & \textbf{92.93 }& 89.49 & 71.08       \\
                      &                      & ShallowCNN & \textbf{99.72} & 92.62 & 67.05       \\
\cline{2-6}
                      & \multirow{3}{*}{MI}  & EEGNet     & 67.49 & \textbf{92.02} & 92.00       \\
                      &                      & DeepCNN    & 72.58 & \textbf{77.30 }& 63.73       \\
                      &                      & ShallowCNN & 61.02 & \textbf{81.90} & 70.89       \\
\midrule
\multirow{6}{*}{PGD}  & \multirow{3}{*}{ERN} & EEGNet     & 94.03 & \textbf{96.92} & 94.38       \\
                      &                      & DeepCNN    & \textbf{97.74} & 89.65 & 72.37       \\
                      &                      & ShallowCNN & \textbf{99.81} & 92.64 & 71.01       \\
\cline{2-6}
                      & \multirow{3}{*}{MI}  & EEGNet     & 65.80 & 95.22 & \textbf{96.88}       \\
                      &                      & DeepCNN    & \textbf{88.08} & 55.53 & 74.78       \\
                      &                      & ShallowCNN & \textbf{91.78} & 87.97 & 87.73       \\
\midrule
\multirow{6}{*}{CW}   & \multirow{3}{*}{ERN} & EEGNet     & 70.40 & 82.17 & \textbf{85.11  }     \\
                      &                      & DeepCNN    & 77.48 & 90.11 & \textbf{90.21}       \\
                      &                      & ShallowCNN & 74.85 & 90.96 &\textbf{ 91.99 }      \\
\cline{2-6}
                      & \multirow{3}{*}{MI}  & EEGNet     & 55.76 & \textbf{71.73} & 69.36       \\
                      &                      & DeepCNN    & 63.08 & 79.80 & \textbf{81.82   }    \\
                      &                      & ShallowCNN & 58.22 & 77.92 & \textbf{80.08  }     \\
\bottomrule
\end{tabular}
\end{table}

Optimization-based white-box attack (CW) was more difficult to detect than gradient-based white-box attacks (FGSM and PGD). Almost all classifiers had significantly lower performance on CW attacks, and the best AUC scores of the eight evaluated adversarial detectors on CW attacks were also lower. The performance differences of these classifiers on gradient-based and optimization-based attacks may reflect, to some extent, the different characteristics of these adversarial examples.

\begin{figure}
\centering
\subfigure[]{
\begin{minipage}[b]{0.4\textwidth}
\includegraphics[width=1\textwidth]{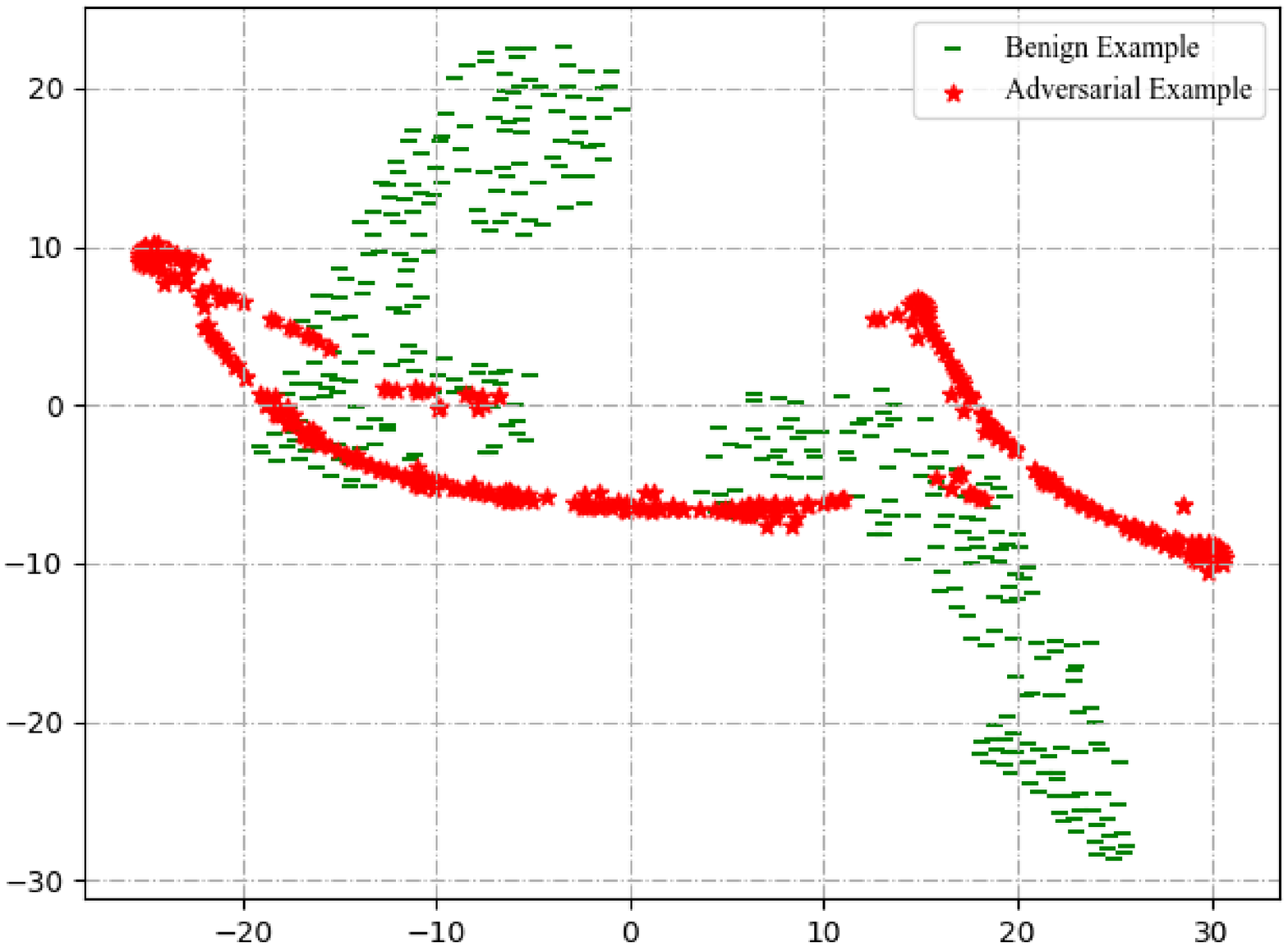}\vspace{0mm}
\end{minipage}
}\\
\subfigure[]{
\begin{minipage}[b]{0.4\textwidth}
\includegraphics[width=1\textwidth]{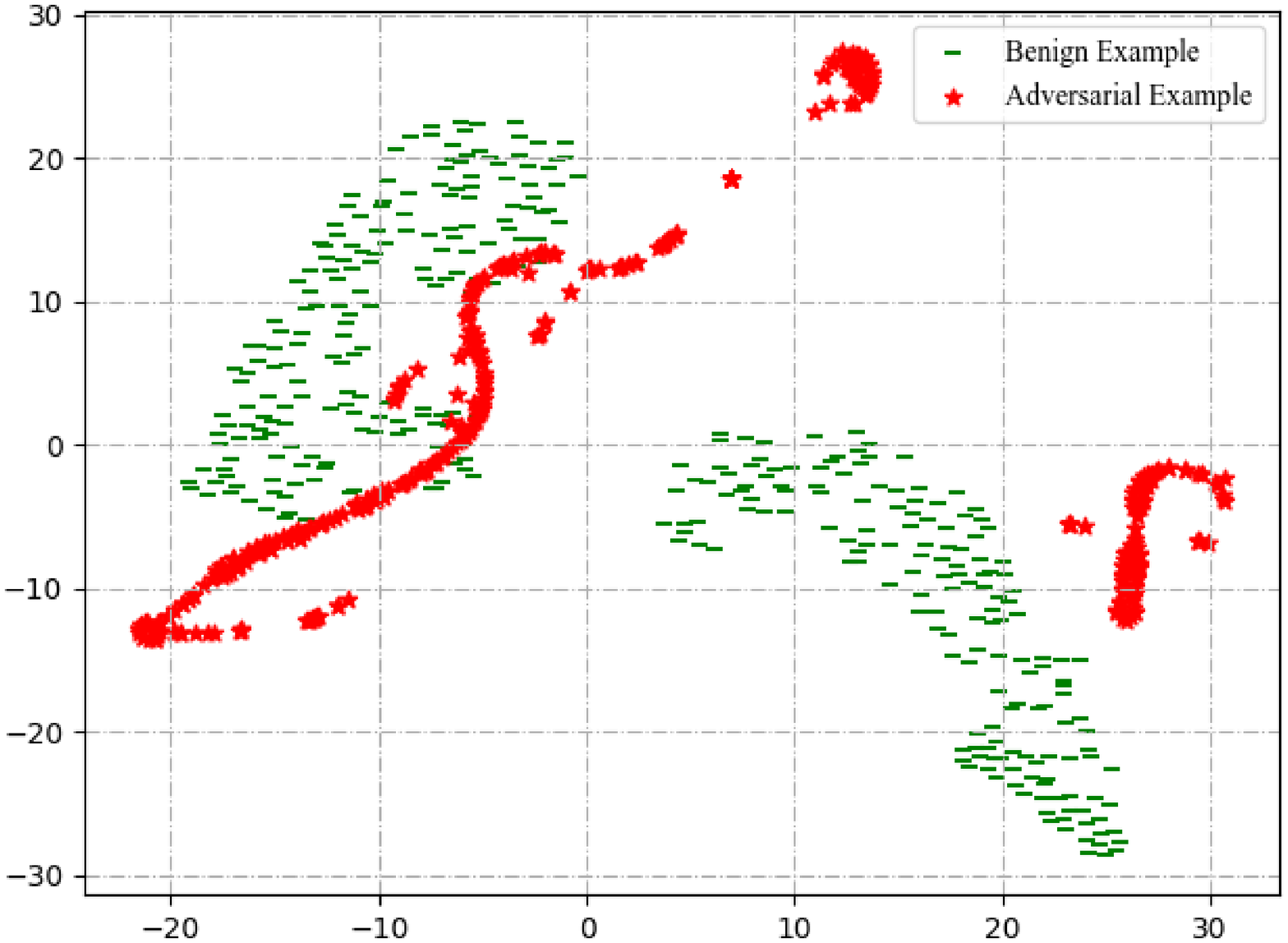}
\end{minipage}}
\\
\subfigure[]{
\begin{minipage}[b]{0.4\textwidth}
\includegraphics[width=1\textwidth]{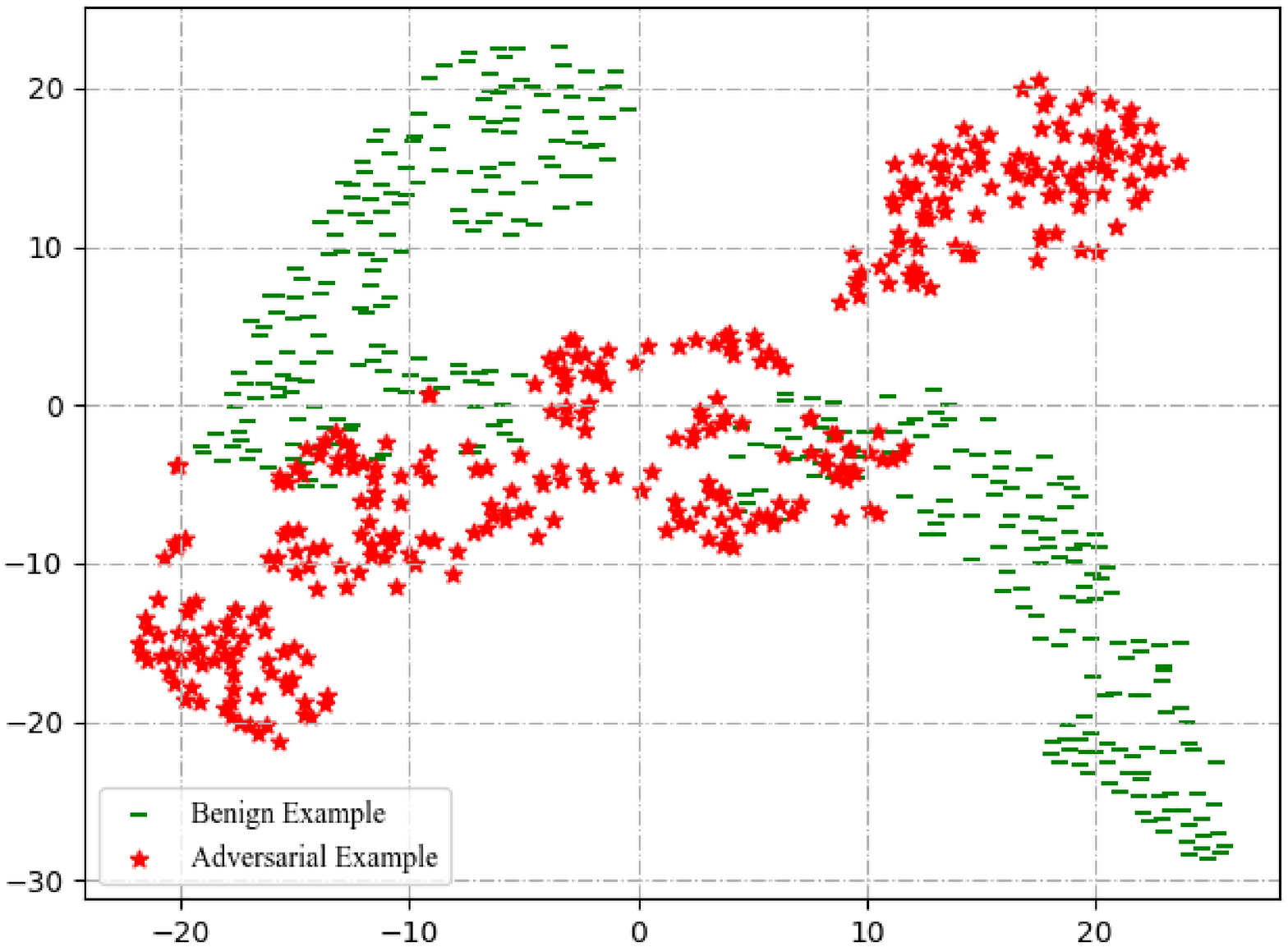}
\end{minipage}
}\caption{$t$-SNE visualization of benign and corresponding adversarial MI EEG epochs in the output space of EEGNet. (a) FGSM; (b) PGD; and, (c) CW.}\label{vi}
\end{figure}

All detectors can distinguish adversarial examples from benign ones to a certain extent, indicating that there are differences between adversarial examples and benign examples. Meanwhile, the performances of the same adversarial detector under different attacks (gradient-based adversarial attack and optimization-based adversarial attack) were different, indicating that there may exist some distribution differences among different adversarial examples. Fig.~\ref{vi} shows $t$-SNE visualizations of distributions of a batch of benign and corresponding adversarial MI EEG epochs in the output space of EEGNet. There are clear differences between the distributions of adversarial and benign examples.

\subsection{Detection of Black-box Attacks}

Table~\ref{tab:black-dec} shows the AUC scores of various detectors under black-box attacks. Compared with white-box attacks, the detection of black-box attacks was more challenging. It should be noted that the detection performance degradation of black-box attacks may not be attributed to the detector itself alone, but may also be because that the adversarial examples generated by black-box attacks were themselves less aggressive and thus their adversarial artifacts were less obvious.

\begin{table}
\centering
\caption{AUC scores (\%) for different adversarial detection approaches under black-box attacks.}
\label{tab:black-dec}
\begin{tabular}{c|c|c|ccc}
\toprule
Dataset              & Target Model                & Substitute  & \makebox[0.01\textwidth][c]  {BU}    & \makebox[0.01\textwidth][c] {LID}            & \makebox[0.01\textwidth][c] MD$_{\max}$  \\
\midrule
\multirow{9}{*}{ERN} & \multirow{3}{*}{EEGNet}     & EEGNet           & 61.10 & \textbf{77.08} & 50.55       \\
                     &                             & DeepCNN          & 59.78 & \textbf{76.11} & 51.53       \\
                     &                             & ShallowCNN       & 51.58 & \textbf{59.05} & 51.08       \\
\cline{2-6}
                     & \multirow{3}{*}{DeepCNN}    & EEGNet           & 48.27 & \textbf{76.01} & 49.90       \\
                     &                             & DeepCNN          & 53.91 & \textbf{78.68} & 51.40       \\
                     &                             & ShallowCNN       & 48.65 & \textbf{57.26} & 51.19       \\
\cline{2-6}
                     & \multirow{3}{*}{ShallowCNN} & EEGNet           & 56.86 & \textbf{63.22} & 52.71       \\
                     &                             & DeepCNN          & 55.39 & \textbf{63.14} & 51.45       \\
                     &                             & ShallowCNN       & 56.54 & \textbf{71.23} & 48.48       \\
\midrule
\multirow{9}{*}{MI}  & \multirow{3}{*}{EEGNet}     & EEGNet           & 71.47 & \textbf{86.59} & 80.72       \\
                     &                             & DeepCNN          & 65.02 & \textbf{69.45} & 53.81       \\
                     &                             & ShallowCNN       & 56.30 & \textbf{71.00} & 53.34       \\
\cline{2-6}
                     & \multirow{3}{*}{DeepCNN}    & EEGNet           & 54.18 & \textbf{61.75} & 54.67       \\
                     &                             & DeepCNN          & 56.30 & \textbf{67.16} & 53.33       \\
                     &                             & ShallowCNN       & 56.36 & \textbf{70.83} & 54.78       \\
\cline{2-6}
                     & \multirow{3}{*}{ShallowCNN} & EEGNet           & 46.02 & \textbf{64.93} & 54.08       \\
                     &                             & DeepCNN          & 47.40 & \textbf{70.12} & 54.68       \\
                     &                             & ShallowCNN       & 48.04 & \textbf{74.64} & 55.64       \\
\bottomrule
\end{tabular}
\end{table}

LID consistently achieved the best performance, may be because:
\begin{enumerate}
\item Unlike the MD feature, whose calculations rely on estimated statistics from the training set, the calculation of the LID of the examples in the test set does not depend on the training set. Since the training and test sets came from different subjects, the LID may be computed more accurately.
\item LID is an estimate of the data dimensionality and is not sensitive to the relative position of the class distribution. Although the distribution of adversarial examples generated under black-box attacks may deviate less from the benign example manifold than that of the adversarial examples generated under white-box attacks, the intrinsic dimensionality of black-box adversarial examples is still higher than that of the benign examples.
\end{enumerate}

\subsection{Generalization to Other Attacks}

To evaluate how well the adversarial detection approaches perform under unknown attacks, we trained adversarial detectors on FGSM adversarial examples, and then evaluated them on PGD and CW. The results are shown in Table~\ref{tab:cross-dec}. Adversarial detectors trained on FGSM still retained good detection performances on PGD, but the classification performances of many detectors on CW were greatly reduced. The  performance differences of various detectors on PGD and CW may indicate that adversarial examples crafted by FGSM and PGD, two gradient-based attacks, have very similar properties, whereas adversarial examples crafted by FGSM and CW, one gradient-based attack and one optimization-based attack, are very different.

\begin{table}
\centering
\caption{Generalization of adversarial detection from FGSM attack to unknown attacks. The LR classifier was trained on the features extracted after applying FGSM attack, and then evaluated on PGD and CW.}
\label{tab:cross-dec}
\begin{tabular}{c|c|c|ccc}
\toprule
Test                 & Dataset              & Model      & \makebox[0.01\textwidth][c]  {BU}    & \makebox[0.01\textwidth][c] {LID}            & \makebox[0.01\textwidth][c] MD$_{\max}$       \\
\midrule
\multirow{6}{*}{PGD} & \multirow{3}{*}{ERN} & EEGNet     & 94.28          & \textbf{96.91} & 94.25           \\
                     &                      & DeepCNN    & \textbf{98.40} & 89.76          & 72.65           \\
                     &                      & ShallowCNN & \textbf{99.88} & 92.64          & 70.92           \\
\cline{2-6}
                     & \multirow{3}{*}{MI}  & EEGNet     & 65.45          & 94.93          & \textbf{96.85}  \\
                     &                      & DeepCNN    & \textbf{91.97} & 78.25          & 73.82           \\
                     &                      & ShallowCNN & 75.48          & \textbf{87.97} & 87.56           \\
\midrule
\multirow{6}{*}{CW}  & \multirow{3}{*}{ERN} & EEGNet     & 30.77          & 79.25          & \textbf{83.52}  \\
                     &                      & DeepCNN    & 22.51          & \textbf{90.36} & 89.34           \\
                     &                      & ShallowCNN & 25.22          & 90.92          & \textbf{91.07}  \\
\cline{2-6}
                     & \multirow{3}{*}{MI}  & EEGNet     & 55.51          & 64.58          & \textbf{69.20}  \\
                     &                      & DeepCNN    & 36.79          & 70.56          & \textbf{81.11}  \\
                     &                      & ShallowCNN & 45.50          & 70.86          & \textbf{79.76}  \\
\bottomrule
\end{tabular}
\end{table}

We also tried to detect FGSM adversarial examples generated under our black-box attack scenario by directly using the adversarial detectors trained using adversarial examples crafted by white-box FGSM attacks. The results are shown in Table~\ref{tab:bc-dec}. Compared with the performances in Table~\ref{tab:black-dec}, the AUC scores of many adversarial detectors were significantly lower (below 50\%), indicating that the distribution of adversarial examples generated in white-box and black-box attacks were quite different, even though both were based on FGSM.

\begin{table}\centering
\caption{Generalization of adversarial detection from white-box FGSM attack to black-box attacks. The LR classifier was trained on the features extracted after applying white-box FGSM attack, and then evaluated on black-box attacks.}
\label{tab:bc-dec}
\begin{tabular}{c|c|c|ccc}
\toprule
Dataset              & Target Model                & Substitute & \makebox[0.01\textwidth][c] {BU}    & \makebox[0.01\textwidth][c] {LID}            &\makebox[0.01\textwidth][c] { MD$_{\max}$}  \\
\midrule
\multirow{9}{*}{ERN} & \multirow{3}{*}{EEGNet}     & EEGNet           & 39.26 & \textbf{75.48} & 52.19       \\
                     &                             & DeepCNN          & 39.25 & \textbf{75.52} & 51.61       \\
                     &                             & ShallowCNN       & 47.33 & \textbf{61.91} & 52.42       \\
\cline{2-6}
                     & \multirow{3}{*}{DeepCNN}    & EEGNet           & 41.61 & \textbf{76.07} & 50.77       \\
                     &                             & DeepCNN          & 48.23 & \textbf{78.64} & 50.55       \\
                     &                             & ShallowCNN       & 45.29 & \textbf{57.13} & 52.24       \\
\cline{2-6}
                     & \multirow{3}{*}{ShallowCNN} & EEGNet           & 41.45 & \textbf{62.79} & 55.51       \\
                     &                             & DeepCNN          & 42.66 & \textbf{62.77} & 54.40       \\
                     &                             & ShallowCNN       & 41.72 & \textbf{70.95} & 52.82       \\
\midrule
\multirow{9}{*}{MI}  & \multirow{3}{*}{EEGNet}     & EEGNet           & 71.43 & \textbf{86.92} & 77.29       \\
                     &                             & DeepCNN          & 65.26 & \textbf{68.78} & 53.51       \\
                     &                             & ShallowCNN       & 56.41 & \textbf{70.67} & 53.61       \\
\cline{2-6}
                     & \multirow{3}{*}{DeepCNN}    & EEGNet           & 45.65 & \textbf{59.45} & 56.52       \\
                     &                             & DeepCNN          & 44.99 & \textbf{65.57} & 55.23       \\
                     &                             & ShallowCNN       & 44.17 & \textbf{69.45} & 56.70       \\
\cline{2-6}
                     & \multirow{3}{*}{ShallowCNN} & EEGNet           & 51.63 & \textbf{62.40} & 54.62       \\
                     &                             & DeepCNN          & 51.44 & \textbf{67.26} & 56.21       \\
                     &                             & ShallowCNN       & 49.99 & \textbf{73.57} & 54.53       \\
\bottomrule
\end{tabular}
\end{table}

\subsection{Discussions} \label{sect:D}

Our detection results under white-box attacks indicated that, when proactive adversarial defense approaches are adopted, there is no need to require the model to fully defend against white-box attacks with strong attack strength. They can be easily recognized by adversarial detection approaches. Furthermore, requiring the model to defend against adversarial examples in adversarial training degrades its performance on benign examples \cite{madry2017towards}.

Adversarial examples generated under black-box attacks are harder to detect than those generated under white-box attacks, but their attack strength is also weaker. Black-box attacks may be better coped with by proactive defense approaches.

For unknown attacks, almost all adversarial detectors have good generalization performance under the same type of attacks (from FGSM to PGD). When the type of adversarial attacks changes, the effectiveness of the adversarial detector may be significantly reduced. This suggests that adversarial examples generated by the same type of adversarial attacks have high similarity, whereas those by different types of adversarial attacks are quite different.

\section{Conclusions and Future Research} \label{sect:CFR}

This paper extended several adversarial detection approaches from computer vision to EEG-based BCIs. We compared their effectiveness under four attacks, including two gradient-based white-box attacks (FGSM and PGD), an optimization-based white-box attack (CW), and a transferability-based black-box attack, on two EEG datasets using three CNN classifiers. We showed that both white-box and black-box attacks can be detected, and the former are easier to detect.

Our future research will:
\begin{enumerate}
\item Explore more metrics for adversarial detection, which describe adversarial examples from more perspectives and may help better understand them.
\item Design feature combinations for adversarial detection. Different features provide different perspectives in adversarial detection. Some may be complementarily, whose combination could improve the adversarial detection performance.
\item Experiment adversarial detections on more adversarial attacks. The properties of adversarial examples generated by different adversarial attacks are usually different. Testing a wider variety of adversarial examples helps better understand adversarial attacks and the vulnerability of the classifiers.
\item Combine proactive adversarial defense approaches with reactive ones. Reactive approaches are better at defending against large perturbations, whereas proactive approaches are better at defending against small ones. There integration may provide more comprehensive defense.
\end{enumerate}



\begin{thebibliography}{10}
\providecommand{\url}[1]{#1}
\csname url@samestyle\endcsname
\providecommand{\newblock}{\relax}
\providecommand{\bibinfo}[2]{#2}
\providecommand{\BIBentrySTDinterwordspacing}{\spaceskip=0pt\relax}
\providecommand{\BIBentryALTinterwordstretchfactor}{4}
\providecommand{\BIBentryALTinterwordspacing}{\spaceskip=\fontdimen2\font plus
\BIBentryALTinterwordstretchfactor\fontdimen3\font minus
  \fontdimen4\font\relax}
\providecommand{\BIBforeignlanguage}[2]{{%
\expandafter\ifx\csname l@#1\endcsname\relax
\typeout{** WARNING: IEEEtran.bst: No hyphenation pattern has been}%
\typeout{** loaded for the language `#1'. Using the pattern for}%
\typeout{** the default language instead.}%
\else
\language=\csname l@#1\endcsname
\fi
#2}}
\providecommand{\BIBdecl}{\relax}
\BIBdecl

\bibitem{ienca2018brain}
M.~Ienca, P.~Haselager, and E.~J. Emanuel, ``Brain leaks and consumer
  neurotechnology,'' \emph{Nature Biotechnology}, vol.~36, no.~9, pp. 805--810,
  2018.

\bibitem{nicolas2012brain}
L.~F. Nicolas-Alonso and J.~Gomez-Gil, ``Brain computer interfaces, a review,''
  \emph{Sensors}, vol.~12, no.~2, pp. 1211--1279, 2012.

\bibitem{he2016deep}
K.~He, X.~Zhang, S.~Ren, and J.~Sun, ``Deep residual learning for image
  recognition,'' in \emph{Proc. {IEEE} Conf. on Computer Vision and Pattern
  Recognition}, Las Vegas, NV, Jun. 2016, pp. 770--778.

\bibitem{devlin2018bert}
J.~Devlin, M.-W. Chang, K.~Lee, and K.~Toutanova, ``Bert: Pre-training of deep
  bidirectional transformers for language understanding,'' in \emph{Proc. Conf.
  of the North American Chapter of the Association for Computational
  Linguistics}, Minneapolis, Minnesota, Jun. 2018, pp. 4171--4186.

\bibitem{parkhi2015deep}
O.~M. Parkhi, A.~Vedaldi, and A.~Zisserman, ``Deep face recognition,'' in
  \emph{Proc. of the British Machine Vision Conf.}, Swansea, UK, Sep. 2015, pp.
  1--12.

\bibitem{szegedy2013intriguing}
C.~Szegedy, W.~Zaremba, I.~Sutskever, J.~Bruna, D.~Erhan, I.~Goodfellow, and
  R.~Fergus, ``Intriguing properties of neural networks,'' in \emph{Proc.
  Int’l Conf. on Learning Representations}, Banff, Canada, Apr. 2014, pp.
  1--10.

\bibitem{goodfellow2014explaining}
I.~J. Goodfellow, J.~Shlens, and C.~Szegedy, ``Explaining and harnessing
  adversarial examples,'' in \emph{Proc. Int’l Conf. on Learning
  Representations}, SanDiego, CA, May 2015, pp. 1--11.

\bibitem{brown2017adversarial}
T.~B. Brown, D.~Man{\'e}, A.~Roy, M.~Abadi, and J.~Gilmer, ``Adversarial
  patch,'' in \emph{Proc. Int’l Conf. on Neural Information Processing
  Systems}, LongBeach, CA, Dec. 2017.

\bibitem{grosse2016adversarial}
K.~Grosse, N.~Papernot, P.~Manoharan, M.~Backes, and P.~McDaniel, ``Adversarial
  perturbations against deep neural networks for malware classification,''
  \emph{arXiv preprint arXiv:1606.04435}, 2016.

\bibitem{carlini2018audio}
N.~Carlini and D.~Wagner, ``Audio adversarial examples: Targeted attacks on
  speech-to-text,'' in \emph{Proc. {IEEE} Security and Privacy Workshops
  ({SPW})}, San Francisco, CA, May 2018, pp. 1--7.

\bibitem{bar2020vulnerability}
A.~Bar, J.~Lohdefink, N.~Kapoor, S.~J. Varghese, F.~Huger, P.~Schlicht, and
  T.~Fingscheidt, ``The vulnerability of semantic segmentation networks to
  adversarial attacks in autonomous driving: Enhancing extensive environment
  sensing,'' \emph{{IEEE} Signal Processing Magazine}, vol.~38, no.~1, pp.
  42--52, 2020.

\bibitem{zhang2019vulnerability}
X.~Zhang and D.~Wu, ``On the vulnerability of {CNN} classifiers in {EEG}-based
  {BCIs},'' \emph{{IEEE} Trans. on Neural Systems and Rehabilitation
  Engineering}, vol.~27, no.~5, pp. 814--825, 2019.

\bibitem{wu2021adversarial}
D.~Wu, W.~Fang, Y.~Zhang, L.~Yang, X.~Xu, H.~Luo, and X.~Yu, ``Adversarial
  attacks and defenses in physiological computing: A systematic review,''
  \emph{National Science Open}, 2021, in press.

\bibitem{meng2019white}
L.~Meng, C.-T. Lin, T.-P. Jung, and D.~Wu, ``White-box target attack for
  {EEG}-based {BCI} regression problems,'' in \emph{Proc. Int’l Conf. on
  Neural Information Processing}, Sydney, Australia, Dec. 2019, pp. 476--488.

\bibitem{zhang2021tiny}
X.~Zhang, D.~Wu, L.~Ding, H.~Luo, C.-T. Lin, T.-P. Jung, and R.~Chavarriaga,
  ``Tiny noise, big mistakes: Adversarial perturbations induce errors in
  brain--computer interface spellers,'' \emph{National Science Review}, vol.~8,
  no.~4, p. nwaa233, 2020.

\bibitem{liu2021universal}
Z.~Liu, L.~Meng, X.~Zhang, W.~Fang, and D.~Wu, ``Universal adversarial
  perturbations for {CNN} classifiers in {EEG}-based {BCI}s,'' \emph{Journal of
  Neural Engineering}, vol.~18, no.~4, p. 0460a4, 2021.

\bibitem{bian2022ssvep}
R.~Bian, L.~Meng, and D.~Wu, ``{SSVEP}-based brain-computer interfaces are
  vulnerable to square wave attacks,'' \emph{Science China Information
  Sciences}, vol.~65, no.~4, pp. 1--13, 2022.

\bibitem{drwuAP2022}
X.~Jiang, L.~Meng, S.~Li, and D.~Wu, ``Active poisoning: Efficient backdoor
  attacks to transfer learning based {BCIs},'' \emph{Science China Information
  Sciences}, 2022, in press.

\bibitem{li2015multimodal}
Y.~Li, J.~Pan, J.~Long, T.~Yu, F.~Wang, Z.~Yu, and W.~Wu, ``Multimodal {BCIs}:
  Target detection, multidimensional control, and awareness evaluation in
  patients with disorder of consciousness,'' \emph{Proc. of the {IEEE}}, vol.
  104, no.~2, pp. 332--352, 2015.

\bibitem{wu2016driver}
D.~Wu, V.~J. Lawhern, S.~Gordon, B.~J. Lance, and C.-T. Lin, ``Driver
  drowsiness estimation from {EEG} signals using online weighted adaptation
  regularization for regression ({OwARR}),'' \emph{{IEEE} Trans. on Fuzzy
  Systems}, vol.~25, no.~6, pp. 1522--1535, 2016.

\bibitem{RR-2996-RC}
A.~Binnendijk, T.~Marler, and E.~M. Bartels, \emph{Brain-computer-interfaces:
  U.S. military applications and implications, an initial assessment}.\hskip
  1em plus 0.5em minus 0.4em\relax Santa Monica, CA: RAND Corporation, 2020.

\bibitem{aldahdooh2022adversarial}
A.~Aldahdooh, W.~Hamidouche, S.~A. Fezza, and O.~D{\'e}forges, ``Adversarial
  example detection for {DNN} models: A review and experimental comparison,''
  \emph{Artificial Intelligence Review}, vol.~55, no.~6, pp. 4403--4462, 2022.

\bibitem{zhang2019theoretically}
H.~Zhang, Y.~Yu, J.~Jiao, E.~Xing, L.~El~Ghaoui, and M.~Jordan, ``Theoretically
  principled trade-off between robustness and accuracy,'' in \emph{Proc.
  Int’l Conf. on Machine Learning}, Long Beach, CA, Jun. 2019, pp.
  7472--7482.

\bibitem{madry2017towards}
A.~Madry, A.~Makelov, L.~Schmidt, D.~Tsipras, and A.~Vladu, ``Towards deep
  learning models resistant to adversarial attacks,'' in \emph{Proc. Int’l
  Conf. on Learning Representations}, Vancouver, Canada, Apr. 2018, pp. 1--28.

\bibitem{feinman2017detecting}
R.~Feinman, R.~R. Curtin, S.~Shintre, and A.~B. Gardner, ``Detecting
  adversarial samples from artifacts,'' \emph{arXiv preprint arXiv:1703.00410},
  2017.

\bibitem{ma2018characterizing}
X.~Ma, B.~Li, Y.~Wang, S.~M. Erfani, S.~Wijewickrema, G.~Schoenebeck, D.~Song,
  M.~E. Houle, and J.~Bailey, ``Characterizing adversarial subspaces using
  local intrinsic dimensionality,'' \emph{arXiv preprint arXiv:1801.02613},
  2018.

\bibitem{lee2018simple}
K.~Lee, K.~Lee, H.~Lee, and J.~Shin, ``A simple unified framework for detecting
  out-of-distribution samples and adversarial attacks,'' in \emph{Proc.
  Advances in Neural Information Processing Systems}, Montreal, Canada, Dec.
  2018.

\bibitem{carlini2017adversarial}
N.~Carlini and D.~Wagner, ``Adversarial examples are not easily detected:
  Bypassing ten detection methods,'' in \emph{Proc. of the 10th ACM Workshop on
  Artificial Intelligence and Security}, Dallas, TX, Nov. 2017, pp. 3--14.

\bibitem{deng2021libre}
Z.~Deng, X.~Yang, S.~Xu, H.~Su, and J.~Zhu, ``Libre: A practical bayesian
  approach to adversarial detection,'' in \emph{Proc. Int'l Conf. on Computer
  Vision and Pattern Recognition}, Nashville, TN, Jun. 2021, pp. 972--982.

\bibitem{zhang2018detecting}
C.~Zhang, Z.~Ye, Y.~Wang, and Z.~Yang, ``Detecting adversarial perturbations
  with saliency,'' in \emph{Proc. 3rd {IEEE} Int'l Conf. on Signal and Image
  Processing ({ICSIP})}, Shenzhen, China, Jul. 2018, pp. 271--275.

\bibitem{carlini2017towards}
N.~Carlini and D.~Wagner, ``Towards evaluating the robustness of neural
  networks,'' in \emph{Proc. {IEEE} Symposium on Security and Privacy}, San
  Jose, CA, May 2017, pp. 39--57.

\bibitem{papernot2016practical}
N.~Papernot, P.~McDaniel, I.~Goodfellow, S.~Jha, Z.~B. Celik, and A.~Swami,
  ``Practical black-box attacks against deep learning systems using adversarial
  examples,'' in \emph{Proc. ACM Asia Conf. on Computer and Communications
  Security}, Abu Dhabi, United Arab Emirates, Apr. 2017, p.~3.

\bibitem{gal2016dropout}
Y.~Gal and Z.~Ghahramani, ``Dropout as a bayesian approximation: Representing
  model uncertainty in deep learning,'' in \emph{Proc. Int'l Conf. on Machine
  Learning}, New York City, NY, Jun. 2016, pp. 1050--1059.

\bibitem{houle2017local}
M.~E. Houle, ``Local intrinsic dimensionality {I}: An extreme-value-theoretic
  foundation for similarity applications,'' in \emph{Proc. Int'l Conf. on
  Similarity Search and Applications}, Munich, Germany, Oct. 2017, pp. 64--79.

\bibitem{margaux2012objective}
P.~Margaux, M.~Emmanuel, D.~S{\'e}bastien, B.~Olivier, and M.~J{\'e}r{\'e}mie,
  ``Objective and subjective evaluation of online error correction during
  {P300}-based spelling,'' \emph{Advances in Human-Computer Interaction}, vol.
  2012, no.~6, p.~4, 2012.

\bibitem{tangermann2012review}
M.~Tangermann, K.-R. M{\"u}ller, A.~Aertsen, N.~Birbaumer, C.~Braun,
  C.~Brunner, R.~Leeb, C.~Mehring, K.~J. Miller, G.~Mueller-Putz \emph{et~al.},
  ``Review of the {BCI} competition {IV},'' \emph{Frontiers in Neuroscience},
  vol.~6, p.~55, 2012.

\bibitem{lawhern2018eegnet}
V.~J. Lawhern, A.~J. Solon, N.~R. Waytowich, S.~M. Gordon, C.~P. Hung, and
  B.~J. Lance, ``{EEGNet}: A compact convolutional neural network for
  {EEG}-based brain-computer interfaces,'' \emph{Journal of Neural
  Engineering}, vol.~15, no.~5, p. 056013, 2018.

\bibitem{schirrmeister2017deep}
R.~T. Schirrmeister, J.~T. Springenberg, L.~D.~J. Fiederer, M.~Glasstetter,
  K.~Eggensperger, M.~Tangermann, F.~Hutter, W.~Burgard, and T.~Ball, ``Deep
  learning with convolutional neural networks for {EEG} decoding and
  visualization,'' \emph{Human Brain Mapping}, vol.~38, no.~11, pp. 5391--5420,
  2017.

\end{thebibliography}
\end{document}